
\magnification 1200
\hsize 15.9truecm
\vsize 22truecm
\baselineskip 18pt
\pageno=0
\footline={\ifnum \pageno< 1 \else \hss \folio \hss \fi }

\null\vskip 3truecm
\centerline {\bf $C_2$ Toda theory in the reduced WZNW framework}
\vskip 2.5truecm
\centerline {{\sl Z. Bajnok}}
\bigskip
\centerline{Institute for Theoretical Physics}
\centerline{Roland E\"otv\"os University}
\centerline{H-1088 Budapest, Puskin u. 5-7, Hungary}

\bigskip
\vskip 2truecm
\underbar{Abstract}
\bigskip

We consider the $C_2$ Toda theory in the reduced WZNW framework.
Analysing the classical representation space of the symmetry
algebra (which is the corresponding $C_2$ $W$ algebra)
we determine its classical highest weight representations. We quantise
the model promoting only the relevant quantities  to operators.
Using the quantised equation of motion we determine the
selection rules for the $u$ field that corresponds to one of the
Toda fields and give restrictions for its amplitude functions and
for the structure of the Hilbert space of the model.

\bigskip
\vfill
\eject

\centerline{\bf Introduction}
\bigskip
\bigskip

The importance of the Toda models lies in the facts that first
they are exactly integrable, furthermore they
give a realisation of extended conformal symmetries, which are
crucial in the classification of conformal field theories
and so
equally important in the string theory and in the analysis of
statistical physical systems.

Since the discovery of the Toda models their investigation,
 classically as well as
at the quantum level, has attracted a great deal of interest.
Analysing the classical
Toda models Leznov and Savaliev showed that their
equation of motion  can be written as zero
curvature conditions and in this way they are exactly
integrable [1]. Then Bilal and Gervais discovered that the symmetry
algebras of the models are the corresponding $W$ algebras [2],
extensions of the chiral Virasoro algebra by some higher spin
chiral currents [3].
It has been shown recently that Toda theories can be regarded as
constrained WZNW models [4]: imposing certain conformal invariant
constraints the WZNW model reduces to the appropriate Toda model
while its symmetry algebra --the Kac-Moody algebra-- becomes the
corresponding $W$ algebra.
The enormous advantage of this picture is that it shows
the relevant variables of the Toda theory explicitly.

Although there have been different approaches to quantise the Toda models
[5-8], much of these efforts concentrated only on the symmetry algebra using
free field constructions.
Concretely in the $C_2$ case Kausch and Watts have shown that
the chiral symmetry algebra is nothing but the commutant of the
screening charges [5]. Using this fact they constructed a free field
representation for the $C_2$ (or $B_2$) $W$ algebra  and
analysed the representation theory of the algebra via
determinant formulae.

The first work in the field of investigating Toda models as
reduced WZNW theories was done in ref. [9] for the simplest case,
the Liouville theory.
Recently (in collaboration with L. Palla and G.
Tak\'acs) we have carried out the analysis for the $A_2$ Toda
theory [10]. We follow this work here with the next simplest
case, $C_2$.
The  interest of this model lies in the non simply-laced nature of the
corresponding Lie algebra. It is an important task to check that the
 methods used, and results obtained for simply-laced algebras can
be generalised for non simply-laced theories.
 Let us remark that the $B_2$ Toda model is
equivalent to the $C_2$ one and even the corresponding $W$ algebras
are the same. However, we work with the $C_2$ model since
the computations in the WZNW description are simpler for this.
(Although the WZNW model depends on this choice the resulting
Toda model is independent of it.)

In this paper we try to make a deeper investigation of the $C_2$ Toda
model using the very natural WZNW framework. As a first step we
compute the defining relations of the
classical $C_2$ $W$ algebra. Then the relevant
degrees of freedom of the Toda
model are identified as the symmetry
generators and the exponential form of one of the Toda fields ($u$).
These variables are not independent since $u$ satisfies the classical
equation of motion.
 With the help of these quantities the classical
representation space of the symmetry algebra is analysed on the solution
space of the classical equation of motion. Characterising
the $W$ orbits by their monodromy matrices we determine the
classical highest weight representations (h.w.r.).
In the quantum case -contrary to other approaches- we promote
only the relevant variables to
operators, which act on a Hilbert space that is the direct sum of
h.w.r. spaces.
The quantum Toda theory is  investigated in three steps. First we
derive the quantum equation of motion - that is the normal
ordered analogue of the classical one - from its covariance.
Then  using this equation we determine the selection rules for
the $u$ field. This gives restrictions on the structure of the
 Hilbert space of the model.
 At last we look for restrictions on the amplitude functions of $u$
by studying the locality requirement for the four point functions.

The paper is organised as follows : In section 1
the advantages of the reduced WZNW description are summarised. Using these
results in section 2 we
describe the classical representation space of the symmetry
algebra. Section 3 deals with the quantum version of the model.
In App. A we show the details of the analysis of the classical
h.w. type solutions. App. B illustrates how we
determine the quantum equation of motion and in App. C we  give
the differential equation that the four point function has
to satisfy.

\vfill
\eject

\vskip 1truecm
\centerline { \bf 1. Classical $C_2$ Toda theory }
\bigskip
\bigskip

The classical $C_2 $ Toda theory is a field theory of two
periodic scalar fields \break
 $ \Phi ^j(x^0,x^1)=\Phi ^j(x^0,x^1+2\pi ); j=1,2$
in two dimensions with exponential interaction:
 $$ L= \sum\limits_{i,j=1}^2
{1\over 2\vert \alpha_i\vert^2}K_{ij}\partial_+\Phi^i\partial_-\Phi^j-
2\sum\limits_{i=1}^2\exp [{1\over 2}\sum\limits_{j=1}^2 K_{ij}\Phi^j]
 \eqno(1.1) $$
 Here  $ x^{\pm}={1\over 2}(x^0\pm x^1)$ are light cone coordinates,
$K_{ij}$ denotes the
Cartan matrix   and $\alpha_i$s denote
the simple roots of the $C_2 $ algebra. The length of the long root of
the algebra is 2.
 The equations of motion are:
$$\partial _+\partial _-\Phi ^1+2e^{\Phi ^1-\Phi ^2}=0$$
$$\partial _+\partial _-\Phi ^2+2e^{\Phi ^2-{1\over 2}\Phi ^1}=
0\eqno(1.2)$$
The model is conformally invariant since the improved,
Feigin-Fuchs type energy momentum tensor  is traceless.
However  the model is invariant  not
only under the Virasoro algebra  but the   $C_2 $
$W$ algebra, too. This algebra, an extension of the
Virasoro algebra, is generated
by  the energy momentum tensor $L(x^+)=W_2(x^+)$ and a spin 4
 chiral current $W(x^+)=W_4(x^+)$.
It is known for a long time that this model  is soluble.
The exact solution of the
Toda models has been found by Leznov and Savaliev in [1].

Another important step was when J. Balog and his
collaborators [4]  showed
that the Toda models can be regarded as
constrained WZNW  models modulo
gauge transformations generated by appropriate conformal
 invariant constraints.
The main advantages of this framework  can be summarised
in the following
points:

 {(i)   The symmetry algebra
- that is the appropriate $W$ algebra -
       can be computed easily : It is the algebra generated
       by the gauge invariant
       polynomials of the constrained currents and their
       derivatives with respect to
       the Dirac bracket. The action of this algebra on the phase space of
        the theory, which is
       the space of the constrained currents, can be
       implemented by certain $KM $
       transformations.}

(ii)  It indicates the relevant degrees of freedom.

(iii) The solutions  of the Toda model can be obtained by reduction from the
       WZNW   solution.

Focusing on the $ C_2 $ case this means that the Toda model can be obtained
by reduction of the $ Sp (2,\bf R)$    WZNW  model. Let us consider now
what  the advantages  mentioned above  mean in this concrete case, when
we work in the so called h.w. gauge:

(i)    The defining relations of the classical $C_2 $ $ W $  algebra are:

$$ \{W_2(x),W_2(y)\}=(W_2(x)+W_2(y))\delta ^{'}(x-y)-5\delta^{'''}(x-y)$$
$$\{W_2(x),W_4(y)\}=-W_4^{'}(y)\delta(x-y)+4W_4(y)\delta^{'}(x-y)$$
$$ \{W_4(x),W_4(y)\}={1\over 2}\sum_{i=0}^2(F_{2i+1}(x)+F_{2i+1}(y))
                                          \delta^{(2i+1)}(x-y)
\eqno (1.3) $$
where
$$ F_5={7\over 25}W_2 \ ; \qquad F_3=-{3\over 5}W_4-{14\over 25}W_2^{''}
                                  -{49\over 125}W_2^2  $$
$$ F_1={14\over 25}W_4W_2+{2\over 5}W_2^{''}+{72\over 625}W_2^3
       +{59\over 125}W_2W_2^{''}+{293\over 500}(W_2^{'})^2
       +{17\over 50}W_2^{''''}  $$

(ii)  The fundamental and proper variables
      of the Toda theory are the generators of
       the symmetry algebra and the lower right
        corner element $u$ of $g$. $g$ is
       the general WZNW  solution whose currents
       satisfy the constraints. It can be
       constructed from the $W$  and $u$.

$$g=\pmatrix{D_4^+D_4^-u & D_4^+D_3^-u & D_4^+D_2^-u & D_4^+ u \cr
             D_3^+D_4^-u & D_3^+D_3^-u & D_3^+D_2^-u & D_3^+ u \cr
             D_2^+D_4^-u & D_2^+D_3^-u & D_2^+D_2^-u & D_2^+ u \cr
                  D_4^-u &      D_3^-u &      D_2^-u &       u }
        \eqno(1.4)$$
        where
$$D_2^{\pm}=-\partial_{\pm} \quad ;\qquad
D_3^{\pm}=-\partial_{\pm}^2+{3\over 10}W_2(x^{\pm}) $$
$$ D_4^{\pm}=-\partial_{\pm}^3+{7\over 10}W_2(x^{\pm})\partial_{\pm}
              +{3\over 10}\partial_{\pm}W_2(x^{\pm}) $$
Here $W_2(x^+)$ and $W_2(x^-)$ denote the left moving and the
        right moving conformal generators, respectively.
        One of the Toda fields now can be written as
    $u=exp\{-{1\over 2} \Phi^2\} $. The other
        one is given with the help of the $2 \times 2$
   lower right sub-determinant of $g$.
        Since the symmetry transformations can be implemented by KM
        transformations their action on $g$ (and thus on $u$ )
   is explicitly given. $u$ turns
        out to be a $W$ primary field, since
$$\eqalign{\delta u =& a_1u^{'}-{3\over 2}a_1^{'}u + \cr
                     &a_2(-u^{'''}+{41\over 50}W_2u^{'}+
           {27\over 100}W_2^{'}u)+a_2^{'}({1\over 2}u^{''}
           -{23\over 100}W_2 u)-{1\over 5}a_2^{''}u^{'}
           +{1\over 20}a_2^{'''}u \cr } \eqno(1.5) $$
where $a_1$ and $a_2$ are infinitesimal functions parameterising
the pure conformal and pure $W$ transformation, respectively [9].
    We distinguish two kinds of Toda solutions.
   If $u$ has no zeroes
         the solution is called regular, and it can be
    expressed in terms of the original
         Toda  variables, the fields $\Phi^i$. In the opposite
     case the solution is called singular,
         and we really need the $u$ field to express them.
 We notice that on this level they are allowed solutions since
     the $W$ densities still remain regular.

(iii)   Since $g$ is the general solution of the WZNW model,
       $u$ must be of the
         following form:
$$ u(x^+,x^-)=\sum_i\psi_i(x^+)\chi_i(x^-)  \eqno(1.6) $$
         As the currents of $g$
    satisfy the constraints, the $\psi_i$s satisfy
          a certain differential equation. The same
     holds for their right handed counterparts,
          $\chi_i$s. Finally the $u$ built up from them
     satisfies the same equation:

$$ u^{''''}-W_2 u^{''}- W_2^{'} u{'}+ \bigl ( W_4+{9\over 100}(W_2)^2
     -{3\over 10}W_2^{''}\bigr ) u =0, \eqno (1.7)$$
      which we call the classical equation of motion,
      since  one of the Toda equations is
      an integral of it.
 This can be seen by expressing $u$ and the symmetry generators in terms of
      the original Toda fields. ( The other Toda equation
      is equivalent to the definition of $\Phi ^1$ with the help of
      $2\times 2$ lower right sub-determinant of $g$.)
          Here the prime means derivative with respect to $x^+$ .
   The boundary conditions of (1.7) are such that at some initial
   point $(x^+_0, x^-_0)$  $g$, built up from $u$, is an element of the
group. Then the evolution of the system, governed by (1.7), will
ensure that $g$ remains in the group.
       A similar equation and boundary condition hold for the right
 moving variables.

\vskip 1cm
\centerline
{\bf 2. The classical representations of the $C_2$ $ W $ algebra }
\bigskip

To find the classical highest weight representations for the $W$
algebra we shall follow the same procedure as we did in ref.[10].
Since the $W$ algebra preserves the form of the constrained
current the transformed $u$ field will satisfy the transformed
equation of motion. This implies that  the corresponding $\Phi$ fields
satisfy the same Toda equation. So the symmetry transformations map
every classical solution into another one. In other words the action
of the symmetry algebra can be represented on the solution space
of the Toda equation.
The solutions connected by $W$  transformations form the so-called
 $W$ orbit.

We are looking for gauge invariant quantities to parameterise these orbits
or representations.
The freedom that only the $u$ field must be periodic, and not the
$\psi $ and $\chi $ fields in(1.5), is coded in the monodromy
matrix:
$$\psi_k(z+2\pi)=M_{kl}\psi_l(z) , \eqno(2.1)$$
(and similarly for $\chi$). The two monodromy matrices are not
independent and they must be chosen correctly to ensure the periodicity
of $u$.
 The monodromy matrices are gauge invariant since the $W$ transformations
 act linearly on the $\psi$ fields, they transform exactly the same way
as $u$ in (1.5).
As suggested above they can be used to parameterise the
representations.
We are interested in the classical highest weight representations
of the $W$ algebra. These are the generalisations of the analogue
quantum representations since we require the following:
The existence of a solution - a highest weight vector - on which
$L$ and $W$ are constant, and for which the total energy $\int W_2 $
 is a minimum on the orbit.

In the rest of this chapter we outline how to find these highest
weight  representations. In three
typical cases we give  explicit results.

We consider diagonalisable monodromy matrices only, since all other are
unphysical in the sense that they never lead to constant $W$ densities.
First we remark that constant $W$ densities are necessary to obtain
h.w. representations ( see appendix for the details). Furthermore it
is not to difficult to show that using constant densities the solutions
of (1.7), which is now a linear differential equation with constant
coefficients, exhaust the cases of diagonalisable monodromy
matrices.

We classify the monodromy matrices by their eigenvalues. From
each class we take a representative, and imposing it as a boundary
condition (2.1), we can determine the linearly independent solutions
of (1.6). These can be used to compute the $W$ densities. If these
densities are periodic and non-singular one has to check
whether the representation obtained is h.w. or not.
This is done by iterating the $W$ transformations. The calculation
is completely analogous to the one explained in [10].

Let us consider the individual cases. If the monodromy matrix has four
real eigenvalues it can be written as
$$ M=diag(e^{\Lambda \pi},e^{-\mu \pi},e^{\mu \pi},e^{-\Lambda \pi})
\qquad \Lambda\not=\mu \eqno(2.2) $$
with $\Lambda$  and $\mu$ arbitrary positive parameters, and let
$\mu > \Lambda $ for convenience.
The corresponding solutions:
$$\eqalign{&\psi_1=N_1e^{\Lambda x^+} \ ;
\quad \psi_4=N_4e^{-\Lambda x^+}\cr
&\psi_2=N_2e^{-\mu x^+} \ ; \quad   \psi_3=N_3e^{\mu x^+}\cr }  $$
where
$$ N_1N_4=(2\Lambda (\mu^2-\Lambda ^2))^{-1} \ ; \qquad
    N_2N_3=(2\mu (\mu^2-\Lambda ^2))^{-1}  $$
 They are normalised to ensure that the matrix built up from them is an
element of SP(2,{\bf R}).
These solutions give constant $W$ densities  :
$$W_2=\mu^2+\Lambda^2  \ ; \qquad  W_4=-{9\over 100}(\mu^2+\Lambda^2)^2
                 +\mu^2\Lambda^2  $$
In the same manner the monodromy matrix of the right moving
fields can be written with some other parameters
$\tilde \Lambda$ and $\tilde \mu$.
The requirement that $u$ must be periodic
identifies $\Lambda$ with $\tilde \Lambda$
and $\mu$ with $\tilde \mu$. Using (1.5)  $u$ can be
built up from its left moving and
right moving components :
$$u=N_1\tilde N_1e^{\Lambda x^0}+N_2\tilde N_2e^{-\mu x^0}
    +N_3\tilde N_3e^{\mu x^0} +N_4\tilde N_4e^{-\Lambda x^0} $$
If the normalisation constants are all negative or positive then
this is a regular solution in the sense that it never changes
its sign, so it can be expressed by the original Toda variables.
In the opposite case the solution may be singular.
This representation can be shown to be classically h.w. for all
possible $\Lambda$ and $\mu$ .

The monodromy matrix that has two real eigenvalues and a complex
conjugate pair can be written in the following form:
$$M=\pmatrix{e^{\Lambda \pi } &      0      &   0    &     0   \cr
            0 & \cos(\pi \rho)  & \sin(\pi \rho) & 0 \cr
            0 & -\sin(\pi \rho) & \cos(\pi \rho)   & 0 \cr
            0 &     0   &    0    &   e^{-\Lambda \pi  }
                                             }\eqno(2.3)$$
where $\Lambda $ and $\rho $  are positive parameters.
Since the monodromy matrix is periodic in $\rho $ only
the $0<\rho < 2 $ region is relevant. The $\psi_i $s which
correspond to this monodromy matrix  are:
$$\eqalign{&\psi_1=N_1e^{-\Lambda x^+} \ ; \quad
\psi_4=N_4e^{\Lambda x^+} \cr
\psi_2&=N_2\sin(\rho x^+)  \ ; \quad  \psi_3=N_2\cos(\rho x^+)\cr }  $$
$$ N_1N_4=(2\Lambda (\rho^2+\Lambda ^2))^{-1} \ ; \quad
    N_2=(\rho (\rho^2+\Lambda ^2))^{-{1\over 2}}  $$
They yield constant $W$ densities:
$$W_2=\Lambda^2-\rho ^2 \ ; \quad  W_4=-{9\over 100}(\Lambda^2-\rho ^2)^2
                 -\rho^2\Lambda^2  $$
We require a similar form for the monodromy matrix of the right
moving variables except for the changing $\Lambda \mapsto \tilde \Lambda$
and $\rho \mapsto \tilde \rho $. The $u$ can be obtained from the $\psi_i$s
and their counterparts:
$$ u=N_1\tilde N_1e^{\Lambda x^0}+N_4\tilde N_4e^{-\Lambda x^0}
+N_2\tilde N_2\cos \bigl({{(\rho-\tilde \rho)}\over 2 }x^0
      +{{(\rho+\tilde \rho )}\over 2 }x^1\bigr)$$
The periodicity of $u$ connects $\rho$ to $\tilde \rho $, namely
$\rho+\tilde \rho =2M $ must hold, where $M$ is an integer.
 It goes without saying
that this solution is a singular one. From the analysis of
the stability condition we conclude that the representation is
h.w. only for $\rho < {1\over 2}, \ \Lambda^2-\rho^2 > -{5\over 2}$.

 Finally let us consider the case when the monodromy matrix has
no real eigenvalues:
$$M=\pmatrix{  \cos(\pi \nu)  & 0  & 0 &-  \sin(\pi \nu )\cr
            0 & \cos(\pi \rho)  & \sin(\pi \rho) & 0 \cr
            0 & -\sin(\pi \rho) & \cos(\pi \rho)   & 0 \cr
      \sin(\pi \nu) &  0   & 0  & \cos(\pi \nu)
                                                    }\eqno(2.4)$$
where again let $\nu >\rho > 0 $ for convenience.
The solutions which satisfy the quasi periodicity conditions are:
$$\eqalign{&\psi_1=-N_1\sin(\nu x^+) \ ; \quad
  \psi_4=N_1\cos(\nu x^+) \cr
& \psi_2=N_2\sin(\rho x^+)  \ ; \quad  \psi_3=N_2\cos(\rho x^+)\cr }  $$
where
$$ N_1=(\nu (\nu ^2-\rho^2))^{-{1\over 2}} \ ; \quad
    N_2=(\rho (\nu ^2-\rho^2))^{-{1\over 2}}  $$
They give constant $W$ densities again:
$$W_2=-\nu^2-\rho ^2  \ ; \quad W_4=-{9\over 100}(\nu^2+\rho ^2)^2
                 +\rho^2\nu^2  $$
Similarly to the previous case the right moving monodromy matrix
is supposed to have the same form but with  $\nu \mapsto \tilde \nu$
and $\rho \mapsto \tilde \rho $. In the usual way the field $u$ can be
written as:
$$u=N_1\tilde N_1\cos\bigl({{(\nu -\tilde \nu )}\over 2 } x^0
      +{{(\nu +\tilde \nu )}\over 2 }  x^1\bigr)
+N_2\tilde N_2\cos\bigl({{(\rho-\tilde \rho)}\over 2 } x^0
      +{{(\rho+\tilde \rho )}\over 2 } x^1\bigr)            $$
Periodicity links $\nu $ to $\tilde \nu$ and $\rho $ to $\tilde
\rho $ as $\nu+\tilde \nu=2K \ ;\ \  \rho+\tilde \rho=2L $ with $K,L$
integers.
As we outline in appendix A this representation is h.w.
only for the range of the parameters
 $\rho <{1\over 2}\ ,\nu >{1\over 2}\ ,\nu-\rho<1$.

This sector contains the classical $SL_2$ invariant vacuum
 as its boundary point. Since
 this solution is singular in the usual
sense, it can be described only with the help of the field $u$,
 we hope to quantise the model using the variable $u$.
 (In the quantum case we are interested at least
in those quantum h.w. representations
 that contain the $SL_2$ invariant vacuum.)
We remark that in the $A_1$ case the classical vacuum,  as here,
is on the boundary of the allowed region, however in the $A_2$ case
the vacuum does not belong to the classically h.w. representations.
The origin of this difference is not clear yet, although it may indicate
that the quantum theory will have a semiclassical limit.
Neither the $A_1$ nor the $A_2$ quantum theory has a semiclassical limit.

\vskip 1truecm
\centerline{ \bf The quantum $C_2$  Toda theory }
\bigskip

We use the WZNW framework to quantise the Toda theory.
 The reasons why we do this be summarised in the following:

On the one hand we would like to describe not only the regular sector of
the classical TT but also the singular one, which can be parameterised only
by the $u$ field.

On the other hand in the quantum case we are interested in those h.w.r.
which  contain the $SL_2$ invariant vacuum. This sector
 could be described classically only by the means of $u$.

Motivated by the classical theory we assume that the relevant degrees
of freedom will become operators in the quantum theory
namely: the generators of the
symmetry algebra, (the quantum $C_2$ $W$ algebra), and $u$.
Actually they are not independent since
$u$ is a primary field with respect to this algebra, and
satisfies the quantum equation of motion that is the normal ordered
 analogue of the classical one. Using this equation,
  which turns out to describe a grade 4 null state, we
could compute the matrix elements of the $u$ field.

We require the symmetry algebra of the model to be the direct
product of the left moving and the right moving  $ C_2$
$W$ algebra. These are generated by  the energy momentum
tensor and the spin 4  current. (The commutation relations
can be found in the literature [11-13].) We assume the
 Hilbert space of the
model to be of the form: ${\cal H}={\cal W}\bigotimes \bar {\cal W}$
where ${\cal W}$ and $\bar {\cal W}$ are built up from h.w.r. spaces.
( From now on we use only the left moving variables,
if it does not lead to  confusion.)
For every h.w. space there exists a h.w. state  $\Bigl\vert
\matrix{h&\bar h\cr
w&\bar w\cr }\Bigr\rangle $, from which the space can be built
up using the Laurent  coefficients of the
generators: $$ W_2(z)=L(z)=\sum_nL_nz^{-n-2}; \qquad
   W_4(z)=W(z)=\sum_nW_nz^{-n-4}  \eqno (3.1) $$

Since the action of these operators  $W^j_n \ (W^2_n=L_n,W^4_n=W_n)$
 on any local field $\phi (z,\bar z)$ can be formulated as:
$$ W_n^j\phi (z,\bar z)=
\oint_z {d\xi \over 2\pi i }(\xi-z)^{n+j+1}W_j(\xi)
\phi(z,\bar z) \eqno(3.2) $$
and the h.w. states correspond to $W$ primary fields in the sense
of [15], the action on the h.w. states is:
$$\matrix{W^j_n\cr \bar W^j_n\cr }\Bigl\vert \matrix{h&\bar h\cr
w&\bar w\cr }\Bigr\rangle =0\qquad n>0\quad j=1,2\eqno(3.3)$$
$$ \matrix{L_0\cr \bar L_0\cr }\Bigl\vert \matrix{h&\bar h\cr
w&\bar w\cr }\Bigr\rangle =\matrix{h\cr \bar h\cr }\Bigl\vert
\matrix{h&\bar h\cr
w&\bar w\cr }\Bigr\rangle \qquad \quad
 \matrix{W_0\cr \bar W_0\cr }\Bigl\vert
 \matrix{h&\bar h\cr
w&\bar w\cr }\Bigr\rangle =\matrix{w\cr \bar w\cr }\Bigl\vert
 \matrix{h&\bar h\cr
w&\bar w\cr }\Bigr\rangle $$

In order to completely describe the model we have to represent the
relevant operators on this Hilbert space. Clearly the action of the
symmetry generators is given so we are concerned with the
representation of the
operator $u$.
We require this operator to be the quantised analogue of the
classical $u$ field. This means that $u$ must be a primary, spin zero,
 periodic field, i.e. :
$$ L_nu(z,\bar z)=\delta_{n,0}\Delta u(z,\bar z) \qquad n\geq 0 $$
$$ W_nu(z,\bar z)=\delta_{n,0}\omega u(z,\bar z) \qquad n\geq 0 $$
where the weights of the field are allowed to differ from their
classical values (1.5) because of the
normal ordering. On quantising the system we use the short distance
OPE to calculate the normal ordered products of the operators.
(Normal ordering is nothing but the subtraction of the singular terms
from the usual OP.)

Furthermore  $u $ must satisfy an equation of motion, that is
the quantum equation of motion, the normal
ordered form of the classical one.
Replacing every term in the classical equation of motion
with a normal ordered term and using the primary nature of $u$ we
find:
$$AL_{-4}u+BW_{-4}u+CL_{-2}^2u+DL_{-3}L_{-1}u+EL_{-2}L_{-1}^2u
                                    +FL_{-1}^4u=0 \eqno(3.4) $$

We remark that although
the classical equation (1.7) contains a cubic term and the normal
ordering is not associative this ambiguity leads only to an
uncertainty of the coefficients. However we have to keep in
mind that, due to the normal ordering,
the coefficients may alter.
These modified values of the parameters ($\Delta, \omega, A, B,
C, D, E, F, $ ) can be determined from the requirement that
the grade 4  null vector defined by eq.(3.4)
 - the quantum equation of motion
 - must transform covariantly under the symmetry algebra.
Denoting the l.h.s. of eq.(3.4) by $\chi $, this means
 that  $\chi $ must be annihilated by $L_n , W_n $
 for $n>0$ or equivalently   $L_1\chi
=W_1\chi =L_2\chi =0$ must hold.
We carry out this analysis in detail in appendix B, and we
find  using an appropriate parameterisation that
$\Delta $ and $\omega $ can be expressed in terms of $Q$ as
$$ \Delta ={1\over 4}(5Q-6) $$
$$ \omega = -{Q\over 8}\Delta
            \sqrt {{{(4Q-5)({8\over Q}-5)({2\over Q}-3)({6\over Q}-7)}
               \over{(226-75Q-{150\over Q})(Q-3)(3Q-7)}}}
      \eqno(3.5) $$
where $Q$ is given by $\ c=(5Q-6)({10\over Q}-6)$.
Since the central charge of the theory is invariant under the
 transformation  $Q \mapsto 2/Q $ we have two kinds of $u$
in a certain model and the $W$ weights of these $u$s are related to
each other by changing every  $Q$ to $2/Q $.
In this parameterisation $Q<0 $ represents the region where $ c>
86+60\sqrt2 $ while $Q>0 $ is the region of $ c<
86-60\sqrt2 $, being in accordance with the defining relations
of the symmetry algebra [11-13]. Consequently $Q\to _-0$
corresponds to the classical limit ($ \Delta \to -{3\over 2}$
, $c \to \infty $).

At this point we are ready to calculate the matrix elements of $u$.
Since the Hilbert space of the model is built up from h.w.r.
spaces it is sufficient to determine the matrix elements between
h.w. states:
$$\Bigl\langle \matrix{H&\bar H\cr \Omega &\bar \Omega\cr }
\Bigr\vert u(z,\bar
z)\Bigl\vert \matrix{h&\bar h\cr
w&\bar w\cr }\Bigr\rangle =G(H,h,\dots )z^{H-h-\Delta }{\bar z}^{\bar
H-\bar h-\bar \Delta }$$
where $G$ may depend on all the parameters describing the
initial and final states.
These amplitude functions are restricted
by the quantum equation
of motion. Sandwiching eq.(3.4) between h.w. states, and using the same
method to compute matrix elements as we did in [10], we get that
$G$ must vanish unless:
$$ \eqalign{ Fy(y+1)(y+2)(y&+3)+E(y+2)(y+3)(y+h)+C(y+h)(y+h+2)+\cr
(D-2E)(y+3)&(y+2h)+(A-2D+2E)(y+3h)+\cr
      B\Biggl(w -3\bigl(
       (&\beta^{-3}-\beta^{-2-1})(y+2h)+\beta^{-2-1}(y+h)(y+2)+\qquad \cr
             &\beta^{-1-1-1}y(y+1)(y+2)\bigr)\cr
     +3\bigl( &\beta^{-2}(y+h)+\beta^{-1-1}y(y+1)\bigr)
            -\beta^{-1}y \Biggr)=0\cr }  \eqno(3.6) $$
with $y=h+\Delta-H$ and the $ \beta $ coefficients are as
defined in appendix B.
It turns out to be very fruitful to introduce the following
reparametrisation for the $W$ weights:
$$ h(a,b)={Q\over 4}(a^2+2ab+2b^2)-{1\over 4}(5Q+{10\over Q}-14)$$
$$\eqalign{ w(a,b)=&\bar B \biggl(
     4Q(Q-3)(27Q-32)b^3(b+2a)-Q(3Q-2)(16Q-27)a^3(a+4b)\cr
   &\qquad +{(Q^2-2)\over
Q}\Bigl(14Q(a^2+2ab+2b^2-6Qa^2b^2)+Q-6+{2\over Q}
   \Bigr)  \biggr) } \eqno(3.7)  $$
where  the $\bar B$ normalisation constant is such that
$\bar B \times B=-{4\over Q}$.
 A similar parameterisation has been used in [5].
In terms of these variables the vacuum corresponds to
 $a_{\rm vac}=(1-2Q^{-1})\ ; b_{\rm vac}=\pm
(1-Q^{-1})$ while $u$ can be described by
$a_u=2-2Q^{-1}$, $b_u=1-Q^{-1}$.

Using this parameterisation we solved (3.6), and found that $u$ has
non-vanishing transition elements if the parameters of the
final and initial states
are linked to each other as:
$$A,\ B= \matrix{a+1,&b\cr
                           a+1,&b-1\cr
                           a-1,&b\cr
            a-1,&b+1\cr } \eqno(3.8) $$
Let us remark that decomposing the tensor product of
two irreps of $SP(2,\bf R)$,
characterised by  Dynkin labels $(a, b)$ and $(1,0)$,
 we get the
 irreducible representations appearing in eq.(3.8).
A similar result was obtained in ref. [14], but for minimal models.

Let us consider now the chiral counterpart of the equation of motion.
Repeating the computation step by step we find the same fusion rules
for $u$ but replacing $a$ with $\bar a$ and $b$ with $\bar b $. Here
$\bar a , \bar b $ parameterise the h.w.r. of the $\bar W $ algebra.
We can relate $a, b$ to $\bar a , \bar b $ by analysing the periodicity
requirement for $u$.
{}From the diagonal transition ($\Bigl\langle \matrix{b+1&a\cr
\bar b+1&\bar a\cr }\Bigr\vert u(z,\bar z)\Bigl\vert \matrix{a&b\cr
\bar a&\bar b\cr }\Bigr\rangle $ etc.)  it follows that
$$ \bar b =b-{{2M}\over Q}\ ; \qquad   \bar a=a+{{2(M+N)}\over Q} $$
where $M, N$ are integers. If $M$ and $N$ were different from zero
 the non diagonal elements would imply among others that
 $Qb$ should be integer, which is excluded as we shall
see later. In this way we conclude that the Hilbert space of the model
may be of the form:
$${\cal H}=\sum\limits_{k,l}{\cal W}_{a_0+k,b_0+l}\otimes \bar {\cal
W}_{a_0+k,b_0+l}\eqno(3.9)$$
where ${\cal W}_{a_0+k,b_0+l}$ is the h.w.r space corresponding to
the following
h.w. state \hfill \break{ $\Bigl\vert
 \matrix{a_0+k&b_0+l\cr
a_0+k&b_0+l\cr }\Bigr\rangle(=\vert a_0+k, b_0+l \rangle $} for short).
This choice is natural in the sense that this Hilbert space may
contain the $SL_2$ invariant vacuum.
{}From now on we will use the following amplitude functions to characterise
$u$ in ${\cal H }$:
$$G_1(a,b)=\langle a+1,b\vert u(1,1)\vert a,b\rangle \  ;\
G_2(a,b)=\langle a+1,b-1\vert u(1,1)\vert a,b\rangle $$
$$G_3(a,b)=\langle a-1,b\vert u(1,1)\vert a,b\rangle  \ ;\
G_4(a,b)=\langle a-1,b+1\vert u(1,1)\vert a,b\rangle $$
{}From the reality of $u$ it follows that:
$$G_3(a,b)=G_1^*(a-1,b)\ ;\qquad G_4(a,b)=G_2^*(a-1,b+1)\eqno(3.10)$$
This means that  there are essentially only two independent amplitude
functions of $u$.

 Let us remark that one can obtain a further restriction for the
amplitude functions analysing the following
automorphism: ${\cal M }\vert a b \rangle =\vert -a-b \rangle $.
This is almost trivial, since due to the special form of the $W$
weights (3.7), ${\cal M} \vert hw\rangle =\vert hw \rangle $ and
$ {\cal M }u(\Delta,\omega){\cal M}^{-1}=u(\Delta,\omega)$.
Applying these results for the matrix elements of $u$ we get:
$$G_1(a,b)=G_1^*(-a-1,-b) \ ; \ G_2(a,b)=G_2^*(-a-1,-b+1)\eqno(3.11)$$

Let us consider the restrictions following from the requirement that
$u$ must be local. These can be established by studying the matrix
elements of $u(z,\bar z)
u(\zeta, \bar \zeta) $. Conformal symmetry restricts this expectation value
to be of the following form:
$$\Bigl\langle \matrix{H&\bar H\cr \Omega &\bar \Omega\cr }\Bigr\vert
 u(z,\bar z)u(\zeta ,\bar \zeta )\Bigl\vert \matrix{h&\bar h\cr
w&\bar w\cr }\Bigr\rangle =(z\zeta )^{\lambda }(\bar z\bar \zeta
)^{\bar \lambda }f(x,\bar x)$$
where $\lambda ={1\over 2}(H-h)-\Delta $ and $x=\zeta/z , \bar x=
\bar \zeta /\bar z $.
The locality of $u$ can be formulated in terms of $f(x,\bar x)$ as
$f(x,\bar x)$ should be invariant under the $x \to x^{-1} , \bar x \to
\bar x^{-1}  $transformation. Since $u$ corresponds to a grade 4 null state
$f(x,\bar x)$ must satisfy a fourth order differential equation.
Solving this d.e. -with the boundary conditions to be described below-
and requiring the symmetry property of the expectation value we obtain
non-linear equations for the amplitude functions. The boundary conditions
can be deduced from the $x\to 0 \ (z\to \infty )$ limit:
$$\langle AB\vert u(z,\bar z)u(\zeta, \bar \zeta)\vert ab\rangle
\rightarrow \sum\limits_{c,d}\langle AB\vert u(z,\bar z)\vert
cd\rangle \langle cd\vert u(\zeta, \bar \zeta)\vert ab\rangle
(1+\dots )=$$
$$=(z\bar z\zeta \bar \zeta)^{\lambda }\sum\limits_{c,d}G(AB;cd)G(cd;ab)
(x\bar x)^{h(c,d)-{1\over 2}(h(A,B)+h(a,b))}(1+\dots )\eqno(3.12)$$
where only those h.w. states give contribution which are allowed by the
selection rules. The dots note polynomials of $x$ and $\bar x $.

To derive   the d.e. we use the usual method to compute matrix elements [10].
After a lengthy but straightforward calculation we determine the d.e., which
is given in appendix C. In order to solve this equation we analyse the
behaviour of the solutions in the vicinity of the singular points
: $x=0 , x=1 , x=\infty $.

At the $x=1  \ (z=\zeta )$ singularity the  indices of
 $f \ (f\sim (1-x)^{\gamma })$
are independent of the  initial and final states and contain
information about
the short distance OPE of $u(z,\bar z) u(\zeta, \bar \zeta)$.
{}From the (C.1) differential equation one can compute the following indices:
$$ \gamma_1=-{1\over 2} (5Q-6) \ ;\quad  \gamma_2=-{1\over 2}
(Q-2) \ ;\quad
  \gamma_3=-{1\over 2} (Q+4) \ ;\quad  \gamma_4={Q\over 2}   $$
They correspond to the presence of operators with conformal weights:
$$ \Delta_1=0 \ ;\quad \Delta_2=2Q-2 \
;\quad \Delta_3=3Q-1 \ ;\quad \Delta_4=3Q-3     $$
It is easy to see that the first is nothing but the identity operator.
The second and the fourth are $W$ primary operators since they
can be parameterised by $a=1-2Q^{-1}\ ,b=2-Q^{-1}$
and $a=3-2Q^{-1} \ ,b=1-Q^{-1} $, respectively.
The operator with conformal weight $\Delta_3$ is a descendant of the
fourth  because its weight differs from $\Delta_4$
by a positive integer.

Let us turn to the study of the $x=0$ singular point.
The indices at this point can be classified by the possible intermediate
states in (3.12).

If there is only one intermediate state, i.e. we are dealing with
 one of the cases
$A,B=a+2,b \ ; \ a+2,b-2 \ ; \ a-2,b \ ; \  a-2,b+2 $, the index is $Q/4$.
Combining this index with the $Q/2$ index of the $x=1$
singularity  we can build up the following trial function:
$$\bigl(x^{-1}(1-x)^2\bigr)^{Q/4}\bigl(\bar x^{-1}(1-\bar
x)^2\bigr)^{Q/4}\eqno(3.13) $$
Using the FORM program [17] we checked that this function solves the
appropriate differential equation. Multiplying it with the corresponding
amplitude functions we can ensure the right $x\to 0$ behaviour.
Since this solution is invariant under the  $x\to x^{-1}, \bar
x\to \bar x^{-1}$ transformation the locality requirement does
not give any restriction for the amplitude functions.

Let us consider the case when there are two intermediate states.
Since there are essentially two independent amplitude functions
it is enough to investigate the following possibilities:
$A,B=a,b+1 $ and $ A,B=a+2,b-1 $.
The corresponding indices are $\pm aQ/4 $ and $\pm (a+2b)Q/4 $.
Motivated by our earlier results [10] we look for the trial
function in the form
$(1-x)^{{Q\over 2}}x^{\rm index}F(\alpha ,\beta ,\gamma ;x)$.
Here $ F $ is the well known hypergeometric function which is
analytic around $x=0$. Its parameters
$\alpha ,\beta ,\gamma  $  can be determined by
requiring the correct asymptotic behaviour around the
singularities, and by demanding the appropriate transformation
properties under $x\to x^{-1}, \bar x \to \bar x^{-1} $.
Combining this form with the corresponding
amplitude functions the trial functions are:
$$\eqalign{\langle a,b+1\vert uu\vert ab\rangle :\quad &\sigma (x)\sigma
(\bar x)\bigl(G_4(a,b)G_1(a-1,b+1)\Psi _a(x)\Psi _a(\bar x)\cr &+
G_1(a,b)G_4(a+1,b)\Psi _{-a}(x)\Psi _{-a}(\bar x)\bigr)\cr }\eqno(3.14)$$
$$\eqalign{\langle a+2,b-1\vert uu\vert ab\rangle :\quad &\sigma(x)\sigma
(\bar x)\bigl(G_1(a,b)G_2(a+1,b)\Psi _{a+2b}(x)\Psi _{a+2b}(\bar x)\cr &+
G_2(a,b)G_1(a+1,b-1)\Psi _{-a-2b}(x)\Psi _{-a-2b}(\bar x)\bigr)\cr }
\eqno(3.15)$$
where $\sigma (x)=\bigl(x^{-1}(1-x)^2\bigr)^{{Q\over 4}}$ and
$$\Psi _a(x)=x^{{Q\over 4}(1+a)}F({Q\over 2},{Q\over 2}(1+a),1
+{Q\over 2}a;x).$$
Using FORM again as in ref.[10] we checked that these solutions solve
the corresponding differential equations.

Let us consider the restrictions given by locality. From the
$x\to x^{-1}$ transformation property of the hypergeometric
functions it follows that:
$$\Psi _a(x)=B_1(a){x^{{Q\over 2}(a+1)}\over (-x)^{{Q\over 2}(a+1)}}
\Psi _a(1/x)+B_2(a){x^{-{Q\over 2}}\over (-x)^{-{Q\over 2}}}
\Psi _{-a}(1/x)\eqno(3.16)$$
where $B_1(a)={\Gamma (1+{Q\over 2}a)\Gamma (-{Q\over 2}a)
\over \Gamma (1-{Q\over 2})\Gamma ({Q\over 2})}$; $B_2(a)=
{\Gamma (1+{Q\over 2}a)\Gamma ({Q\over 2}a)\over \Gamma
({Q\over 2}(a-1)+1)\Gamma ({Q\over 2}(a+1))}$
This implies for the amplitude functions that:
$${G_1(a,b)G_2(a+1,b)\over G_2(a,b)G_1(a+1,b-1)}=\phi (a+2b) $$
$${G_1(a,b)G_4(a+1,a)\over G_4(a,b)G_1(a-1,b+1)}=\phi (a)$$
where $$\phi (a)=-{\Gamma ^2(-{Q\over 2}a)\Gamma
({Q\over 2}(a+1))\Gamma (1+{Q\over 2}(a-1))\over
\Gamma ^2({Q\over 2}a)\Gamma ({Q\over 2}(1-a))\Gamma
 (1-{Q\over 2}(a+1))}={s(a+1)\over s(a-1)}
{\Gamma ^2(-{Q\over 2}a)\Gamma ^2({Q\over 2}(a+1))\over
 \Gamma ^2({Q\over 2}a)\Gamma ^2({Q\over 2}(1-a))} $$
with $s(x)={\rm sin}(\pi {Q\over 2}x)$.
Using these equations one can show that the amplitude functions
must have the following form:
$$\vert G_1(a,b)\vert ^2=  f_2(a,b){\Gamma ({Q\over 2}(a+2b+1))
\Gamma (-{Q\over 2}(a+2b))\Gamma
(-{Q\over 2}a)\Gamma ({Q\over 2}(a+1))\over
 \Gamma (1-{Q\over 2}(a+2b+1))\Gamma (1+{Q\over 2}(a+2b))\Gamma
(1+{Q\over 2}a)\Gamma (1-{Q\over 2}(a+1))} $$
$$\vert G_2(a,b)\vert ^2= f_1(a,b){\Gamma ({Q\over 2}(a+2b))\Gamma
 (-{Q\over 2}(a+2b-1))\Gamma
({Q\over 2}(a+1))\Gamma (-{Q\over 2}a)\over
\Gamma (1-{Q\over 2}(a+2b))\Gamma (1+{Q\over 2}(a+2b-1))\Gamma
(1-{Q\over 2}(a+1))\Gamma (1+{Q\over 2}a)} \eqno(3.17) $$
where $f_1(a,b)$ is invariant under the transformation $a\to a+1$
-essentially independent of $a$- and $f_2(a,b)$ is invariant
under the transformation $a\to a+1, b\to b-1 $ (i.e. $f_2(a,b)$
essentially depends on $a+b$).

 Finally let us consider the diagonal transition. In this case
there are four intermediate states: $ a+1,b \ ;\  a+1,b-1 \ ; \
a-1,b \ ; \ a-1,b+1 $. The corresponding indices around $x=0$
are:
$$\nu_1={Q\over 4}(1-2a-2b) \ ;\ \nu_2={Q\over 4}(1-2b) \ ;\
\nu_3={Q\over 4}(1+2a+2b) \ ;\ \nu_4={Q\over 4}(1+2b) $$
which are the same as those that come from the differential equation.
 In order to define a
periodic $u$ it is necessary to avoid the appearance of
logarithmically singular solutions.
Since our differential equation is of the Fuchs type if we want
to avoid the logarithmic singularities of its solutions we have to
demand that no pairs of the indices differ by an integer.
This is exactly the same restriction that we used to establish
the Hilbert space of the model (3.9).
The next step would be to construct the solutions of the d.e..
However the d.e. is so complicated that we have not succeeded in solving it
in the usual way.
Using a free field representation for the $C_2 W $
algebra [5] we found solutions (with the appropriate asymptotic
behaviour for $x=0$) in terms of triple contour integrals.
Unfortunately we could determine neither the transformation property
of the solutions under $x \mapsto 1/x$ nor the monodromy matrix
corresponding to the singular points.
In this way the $f$ functions remain arbitrary (except for the fact
that they have to satisfy (3.11) since the factors occurring in
(3.17) satisfy it.)
In order to determine them one has to either carry out a complete
analysis of the four point functions (that we have not succeeded in),
 or analyse higher point functions.

\bigskip
\centerline{\bf Conclusions }
\bigskip

In this paper we successfully applied the reduced WZNW framework for
the $C_2$ Toda theory. At the classical level we could generalise the
result obtained previously for $A_2$:
 We derived the classical $C_2 $ W algebra in the
highest weight gauge. Identifying the relevant degrees of freedom
and declaring the connection between them (the classical equation of
motion) we established a framework in which we analysed the classical
representation space of the symmetry algebra.
Parameterising the $W$ orbits by their
monodromy matrices we identified those
which correspond to classically h.w.r.. We showed that,
contrary to the $A_2$ case, the orbit of the classical $SL_2$
invariant vacuum is of the h.w. type and is on the boundary of the
allowed region.

In the quantum case we quantised
the construction above ( which was successful classically ),
in contrast to other
approaches. Promoting only the
relevant variables to operators we required the symmetry algebra to be
the $C_2$ quantum  $W$ algebra [11-13]. We supposed the Hilbert space of
the model to contain only h.w.r. spaces and demanded $u$ to be a
periodic $W$ primary field, which satisfied the quantum equation of
motion that is the normal ordered analogue of the classical one.
Using this q.e.m. we obtained the selection rules for the $u$ field and
partly determined the relevant  amplitude functions, analysing
the locality requirement of the four point functions. This means
that at the quantum level we could not completely generalise  the
result obtained previously. However we should remark that the
problems are purely technical. One can go further
 analysing higher point functions or considering other operators
in the theory appearing for example in the OPE of $u$ with itself.
This operator is nothing but the field which corresponds to the
$B_2$ theory [16]. Representing this field on the same Hilbert space
and considering its four point functions with $u$ simultaneously
a complete analysis
could be made.
This may lead to constraints on the value of the
central charge and on the structure of the Hilbert
space of the model.
\vskip 1truecm

\centerline{\bf Acknowledgements}

\bigskip
I would like to thank L. Palla, Z. Horv\'ath, P. B\'antay for the many
useful discussions and comments. This work was partly supported by
the Sz\'echenyi Istv\'an \"Oszt\"ond\'{\i}j Alap\'{\i}tv\'any.

\vfill
\eject

\centerline { \bf Appendix A }
\bigskip

\centerline {\bf Stability analysis}
\smallskip

In this appendix we investigate the condition for an
orbit to correspond to classical h.w.r. As we defined
earlier, the h.w. solutions -described by the various monodromy matrices-
must give constant $W$ densities ($L(z)=L_0 , W(z)=W_0 $ respectively)
and the total energy of the system has to increase moving along the
orbit. In order to check that the energy does increase we iterate the
$W$ transformations. In the first approximation the variation of the
$W$ densities cancel since $L_0,W_0$ are constant and the $a_i$
transformations functions are periodic:
$$ \delta L=\int_0^{2\pi}(2a_1^{'}L_0-5a_1^{'''}+4a_2^{'}W_0)dz =0$$
$$\eqalign{\delta W=\int_0^{2\pi}\biggl(4&a_1^{'}W_0+a_2{'}\bigl(
 {14\over 25}
L_0W_0+{72\over 625 }L_0^3\bigr)+\cr &a_2{'''}\bigl( -{3\over 5}W_0
-{49\over 125}L_0^2\bigr) +{7\over 25} a_2^{(V)}L_0
-{1\over 20} a_2^{(VII)}\biggr)dz=0 \cr }$$

We remark that in the case of non-constant $W$ densities we can always
choose such $a_1,a_2$, st. $\delta L$ is  positive or negative, which means
that non-constant $W$ densities never give minima or maxima.
Going back to the formula the point described by $L_0, W_0$ turns out
to be a stationary point of the orbit, and it is necessary to iterate further.
Keeping in mind that the $W$ densities are no longer constants
(due to the arbitrary $a_i$ functions in $\delta W_i$) we could write
$\delta \delta L$:
$$ \delta \delta L=\int\limits_0^{2\pi }(a_1^{,}\delta
L+a_2^{,}\delta W)dz $$
which can be rewritten as
$$ \delta \delta L=\int\limits_0^{2\pi }
   \pmatrix { a_1^{,} & a_2^{,} }
   \pmatrix { 2L_0-5{d^2\over dz^2 } & 4W_0 \cr
    4W_0 & {\cal D}   \cr }
   \pmatrix { a_1^{,} \cr a_2^{,} }dz \eqno(A.1) $$
$$ {\cal D}= \bigl( {14\over 25}
L_0W_0+{72\over 625 }L_0^3\bigr) +\bigl( -{3\over 5}W_0
-{49\over 125}L_0^2\bigr){d^2\over dz^2 } +{7\over
25}L_0{d^4\over dz^4 }
-{1\over 20}{d^6\over dz^6 }  $$
where we have dropped the total derivatives. Let us analyse this
matrix in a convenient basis of the form:
$$ \pmatrix { a_1^{,} \cr a_2^{,} }={\bf q}e^{inz} ,\qquad n\not= 0 $$
which span the space of the possible transformations.

For a solution to belong to a  h.w.r. it is sufficient and necessary
that its matrix (in terms of n)
$$ M(L_0,W_0)= \pmatrix { 2L_0+5n^2 & 4W_0 \cr
    4W_0 & {\cal D}(n)
   \cr } $$
$${\cal D}(n)=  \bigl( {14\over 25}
L_0W_0+{72\over 625 }L_0^3\bigr) +\bigl( {3\over 5}W_0
+{49\over 125}L_0^2\bigr)n^2 +{7\over 25}L_0n^4
+{1\over 20}n^6  $$
be positive definite.
This condition can be formulated as the positivity of the
left upper component and the determinant of M. It is evident
that it is sufficient to consider the $n=1$ case. The first requirement
is nothing but the positivity of the energy.
The second can be transformed into the following inequality:
$$\bigl((4a+1)+16b\bigr)\bigl({1\over 4}(a+1)^2-b\bigr)>0 \eqno(A.2)$$
where we parameterised $L_0 $ and $W_0 $ as:
$$L_0=a \qquad W_0=-{9\over 100}a^2+b $$

Fortunately all three cases discussed in chapter 2
are included in this equation
with an appropriate choice of the region of the parameters.
Analysing the inequality one can show that in the first case $a>0, b>0$
-described by the monodromy matrix (2.3) - there is no restriction on the
parameters $\Lambda $ and $\mu$. In such a way the representation is
h.w.r. for all  $\Lambda $ and $\mu$.
The monodromy matrix with two real eigenvalues corresponds to the case
of $a>0, b<0$, in which case the positive definiteness condition
implies that $\rho$
must be smaller then ${1\over 2}$. From the positivity of the energy,
$\Lambda^2-\rho^2>{5\over 2} $ must hold.  The representation is
 h.w. only for these values of the parameters.
 The last case, when $a<0\ , b>0 $, describes the monodromy matrix with no
real eigenvalues. Using both requirements ( for the energy and the
determinant) one can show that $\rho < {1\over 2} \
\nu >{1\over 2} \ \nu-\rho <1 $ is needed.
Since the classical $SL_2$ invariant vacuum, described by
 $ L_0=-{5\over 2} \ , W_0=0$, is on the boundary of the possible region,
 the quantum theory may have a classical limit in contrast to
the $A_2$ case. Finally we notice that the minima obtained are the only
possible minima  since they are the only points of the orbits which give
constant $W$ densities and in this way lead to minima. Clearly the $W$
densities are uniquely determined by the various gauge invariant
monodromy matrices.

\vskip 1cm
\centerline{\bf Appendix B }
\bigskip
\centerline {\bf The covariance of the quantum equation of motion }
\smallskip
The covariance of equation (3.5) ($ \chi $)
 - i.e. $L_n\chi=0 \ , W_n\chi =0 $ for $n>0$
implies that there must exist a null state,$\phi $,
 on grade 3 of the  following form:
$$\phi =\beta^{-3}L_{-3}u+W_{-3}u+\beta^{-2-1}L_{-2}L_{-1}u
                        +\beta^{-1-1-1}L_{-1}^3u=0 \eqno(B.1) $$
{}From the requirement that the generators with positive indices
annihilate $\chi $ and $\phi $ it
follows that there are two independent null states on grade 2 and
grade 1:
$$\beta^{-1} L_{-1}u+W_{-1}u=0  \eqno(B.2) $$
$$\beta^{-2}L_{-2}u+W_{-2}u+\beta^{-1-1}L_{-1}^2u=0, \eqno(B.3) $$
respectively.
Since all these states have to be null states one can compute the
various $\beta $ coefficients, iteratively. We find for the first two
coefficients that
$$\beta^{-1}=-2{\omega \over \Delta }$$
$$\beta^{-2}=2(23-10Q)N  \qquad  \beta^{-1-1}=4(13-{25\over Q})N
 \eqno(B.4)$$
with
$$ N={\omega \over \Delta }{1\over{(4Q-5)({8\over Q}-5)}} $$
and  these conditions fix the $W$ weights of the $u$ operator,
($\Delta , \omega $ ).

Furthermore
 $$\beta^{-3}=-6(20Q^3-17Q^2-116Q+108)M $$
$$\beta^{-2-1}=-24(34Q^2-113Q+82)M \eqno(B.5) $$
$$\beta^{-1-1-1}=-16(226-75Q-{150\over Q})M $$
where
$$ M= {\omega \over \Delta }{1\over{(4Q-5)({8\over
Q}-5)(7Q-6)(3Q-2)}} $$

As a consequence of these results the coefficients of the quantum equation
of motion are:
$$ A= -{2\over Q}(30Q^2-23Q^3+178Q^2-936Q+720)$$
$$ B=-8Q\sqrt{(4Q-5)({8\over Q}-5)({2\over Q}-3)({6\over Q}-7)
            (226-75Q-{150\over Q})(Q-3)(3Q-7)} $$
$$ C=4(16Q-27)(3Q-2)$$
$$ D= -{8\over Q}(27Q^3+37Q^2-356Q+300) \eqno(B.6)$$
$$ E= 16(226-75Q-{150\over Q})$$
$$ F=-{16\over Q}(226-75Q-{150\over Q})$$

\bigskip
\smallskip
\centerline { \bf Appendix C }
\bigskip

\centerline {\bf The differential equation for $f(x,\bar x)$ }
\smallskip
In this appendix we describe the differential equation
which the four point function, $f$, has to satisfy.
 Sandwiching the quantum equation of
motion and using the freedom to deform the contour
in the integral representation of the $W$ generators, (3.2),
 we get the following
equation:
$$\eqalign{\Biggl\{ F{\cal L}_{-1}^{(4)}
+(D-2E)({\cal L}_{-1}^{(1)}-3){\cal L}_{-3}&+
E({\cal L}_{-1}^{(2)}-4{\cal L}_{-1}^{(1)}+6){\cal L}_{-2}+\cr
C({\cal L}_{-2}+2){\cal L}_{-2}+(A-&2D+2E){\cal L}_{-4}+\cr
    B\Bigl({\omega\over {(1-x)^4}}+w+({1\over {(1-x)^3}}-1)
 &(-\beta^{-1}\hat {\cal L}_{-1}^{(1)})+\beta^{-1}{\cal L}_{-1}^{(1)}\cr
+\bigl( -2x-1+{1\over{(1-x)^2}}\bigr)&(-\beta^{-2}\hat{\cal L}_{-2}
                                   -\beta^{-1-1}\hat {\cal L}_{-1}^{(2)})
               +\cr
                                      3\beta^{-2}{\cal L}_{-2}
                        +3\beta^{-1-1}&{\cal L}_{-1}^{(2})+\cr
 \bigl({1\over{(1-x)}}-(x-1)^2-3x\bigr)&\bigl(
                        -(\beta^{-3}-\beta^{-2-1})
\hat {\cal L}_{-3}-\beta^{-2-1}\hat {\cal L}_{-1}^{(1)}\hat {\cal L}_{-2}\cr
           &\hskip 3.5truecm -\beta^{-1-1-1}\hat {\cal L}_{-1}^{(3)}
            \bigr)\cr
          +3(\beta^{-3}-\beta^{-2-1}){\cal L}_{-3}
                    & +3\beta^{-2-1}({\cal L}_{-1}^{(1)}-2){\cal L}_{-2}
           +3\beta^{-1-1-1}{\cal L}_{-1}^{(3)} \Bigr)\Biggr\}f=0
           \cr }\eqno(C.1) $$

where

$$\eqalign{{\cal L}_{-n}=(n-1)\bigl({\Delta\over {(x-1)^n}}+(-1)^nh\bigl)
             & -{1\over{(x-1)^{n-1}}}({\lambda\over x}+ d_x)\cr
        &-(-1)^n\bigl( \lambda(1+{1\over x})+(1-x)
                                          d_x\bigr)\cr } $$

$$\eqalign{\hat {\cal L}_{-n}=(n-1)\bigl({\Delta\over {(1-x)^n}}
                                    +(-1)^n{h\over{x^n}}\bigl)
             & -{1\over{(1-x)^{n-1}}}(\lambda-x d_x)\cr
        &-{{(-1)^n}\over{x^{n-1}}}\bigl( \lambda(1+{1\over x})+(1-x)
                                          d_x\bigr)\cr } $$

and

$${\cal L}_{-1}^{(n)}=\sum_{k=0}^n(-1)^k\lambda^{(n-k)}x^k d_x^k
 \qquad \lambda^{(k)}=\prod_{i=k}^{n-1}(\lambda-i)  $$

$$\hat {\cal L}_{-1}^{(n)}=\sum_{k=0}^n \lambda^{(n-k)}{1\over
{x^k}} d_x^k
 \qquad \lambda^{(k)}=\prod_{i=0}^{k}(\lambda-i)  $$

and $d_x $ means derivative with respect to $x$.

\vskip 2truecm
\centerline{\bf References }

\item{[1]} A.N. Leznov and M.V. Savaliev: {\sl Lett. Math.
Phys.} {\bf 3} (1979) 489; {\sl Comm. Math. Phys.} {\bf 74}
(1980) 111.
\item{[2]} A. Bilal and J.L. Gervais : {\sl  Nucl. Phys.} {\bf
B318} (1989) 579.
\item{[3]}
P. Bouwknegt and K. Schoutens, {\sl Phys. Rep.} {\bf 223} (1993) 183.
\item{[4]}
P. Forg\'acs, A. Wipf, J. Balog, L. Feh\'er and L. O'Raifeartaigh:
{\sl Phys.
Lett.} {\bf B227} (1989) 214; {\bf B244} (1990) 435; {\sl Ann.
Phys. (N. Y.)} {\bf 203} (1990) 76.
\item{[5]}
H.G.Kausch and G.M.T.Watts Int. J. Mod. Phys. A7 4175 (1992),
 {\sl  Nucl. Phys.} {\bf 386} (1992) 166.
\item{[6]} A. Bilal and J.L. Gervais : {\sl  Phys. Lett.} {\bf
B206} (1988) 412, {\sl Nucl. Phys.}
{\bf B314} (1989) 646.
\item{[7]}
 L. Feh\'er and L. O'Raifeartaigh, P. Ruelle, I. Tsutsui and
 A. Wipf: {\sl Phys. Rep.} {\bf 222}, No. 1 (1992) 1-64
\item{[8]}
 P. Mansfield : {\sl  Nucl. Phys.} {\bf B222} (1983) 419.
\item{[9]} J. Balog, L. Palla: {\sl Phys. Lett.} {\bf B274} (1992) 323.
\item{[10]}
Z. Bajnok, L. Palla and G. Tak\'acs  {\sl  Nucl. Phys.} {\bf
B385} (1992) 329.
\item{[11]}
K.-J. Hamada and M. Takao {\sl Phys. Lett.} {\bf B209} (1988) 247.
\item{[12]}
Bouwknegt {\sl Phys. Lett.} {\bf B207} (1988) 295.
\item{[13]}
R. Blumenhagen, M. Flohr, A. Kliem, W. Nahm, A. Recknagel and
R. Varnhagen  {\sl  Nucl. Phys.} {\bf B361} (1991) 255.
\item{[14]}
E. V. Frenkel, V. Kac and M. Wakimoto, {\sl Commun. Math. Phys.}
{\bf 147} (1992) 195.
\item{[15]} A.A. Belavin, A.M. Polyakov and A.B. Zamolodchikov:{\sl
Nucl. Phys.} {\bf B241} (1984) 333.
\item{[16]} Z. Bajnok,  {\sl Phys. Lett.} {\bf B320 } (1994) 36.
\item{[17]} J. Vermaseren: FORM User's guide, Nikef Amsterdam, April 1990.

\vfill
\end